\documentclass[aps,prl,twocolumn,groupedaddress,linenumbers]{revtex4}
\usepackage{graphicx,color}
\usepackage{amsmath,amssymb,latexsym}
\usepackage{mathtools}
\usepackage[english]{babel}
\usepackage{dcolumn}
\usepackage{bm}
\usepackage{float}
\usepackage[normalem]{ulem}

\usepackage{color}

\def\cL{ {\mathcal L} }
\def\cR{ {\mathcal R} }
\begin{document}

\title{Polymers critical point originates Brownian non-Gaussian diffusion}

\author{Sankaran Nampoothiri}
\email{sankaran.nampoothiri@unipd.it}
\affiliation{
Dipartimento di Fisica e Astronomia `G. Galilei' - DFA, Sezione INFN,
Universit\`a di Padova,
Via Marzolo 8, 35131 Padova (PD), Italy
}

\author{Enzo Orlandini}
\email{orlandini@pd.infn.it}
\affiliation{
Dipartimento di Fisica e Astronomia `G. Galilei' - DFA, Sezione INFN,
Universit\`a di Padova,
Via Marzolo 8, 35131 Padova (PD), Italy
}

\author{Flavio Seno}
\email{flavio.seno@unipd.it}
\affiliation{
Dipartimento di Fisica e Astronomia `G. Galilei' - DFA, Sezione INFN,
Universit\`a di Padova,
Via Marzolo 8, 35131 Padova (PD), Italy
}

\author{Fulvio Baldovin}
\email{fulvio.baldovin@unipd.it}
\affiliation{
Dipartimento di Fisica e Astronomia `G. Galilei' - DFA, Sezione INFN,
Universit\`a di Padova,
Via Marzolo 8, 35131 Padova (PD), Italy
}

\date{\today}

\begin{abstract}
We demonstrate that size fluctuations close to polymers critical point
originate the non-Gaussian diffusion of their center of mass. Static
universal exponents $\gamma$ and $\nu$ -- depending on the polymer
topology, on the dimension of the embedding space, and on
equilibrium phase -- concur to determine the potential divergency of a
dynamic response, epitomized by the center of mass kurtosis.
Prospects in experiments and stochastic modeling brought about by this
result are briefly outlined.
\end{abstract}

\maketitle

As a consequence of the central limit theorem,
ordinary diffusive
motion of mesoscopic particles in solution is
characterized by a Gaussian probability density function (PDF) whose 
variance grows linearly over time.
Numerous experiments performed
in complex
contexts~\cite{wang2009,wang2012,toyota2011,chakraborty2020,weeks2000,wagner2017,jeon2016,yamamoto2017,stylianidou2014,parry2014,munder2016,cherstvy2018,li2019,cuetos2018,hapca2008,pastore2021rapid},
while confirming the linear temporal increase of the variance, 
highlight however distinct stages during which the PDF of the random
motion is non-Gaussian. This interesting contingency has been called
``Brownian non-Gaussian diffusion'' and has inspired various
mesoscopic approaches, invoking superposition of
statistics~\cite{beck2003,beck2006,hapca2008,wang2012}, diffusing
diffusivities~\cite{chubynsky2014,chechkin2017,jain2017, tyagi2017,
  miyaguchi2017, sposini2018, sposini2018first,miotto2021length}, 
subordination concepts~\cite{chechkin2017}, continuous time random
walk~\cite{barkai2020}, and diffusion in disordered
environments~\cite{sokolov2021}, but presently few attempts
have been made to establish a microscopic foundation of this
phenomenon~\cite{baldovin2019,hidalgo2020}.  To breach the central
limit theorem~\cite{Feller1968} one possibility is the emergence of
strong correlations; here we demonstrate that the polymers critical
point, separating the dilute to the dense phase in the grand canonical
ensemble~\cite{deGennes1972,deGennes1979,vanderzande1998,madras2013},
indeed originates a Brownian yet non-Gaussian diffusion for the center
of mass (CM) of a polymer in solution.  Prospects in experiments
and stochastic modeling brought about by this result are briefly
outlined.

\begin{center}
\begin{figure}[t]
\includegraphics[width=0.65\columnwidth]{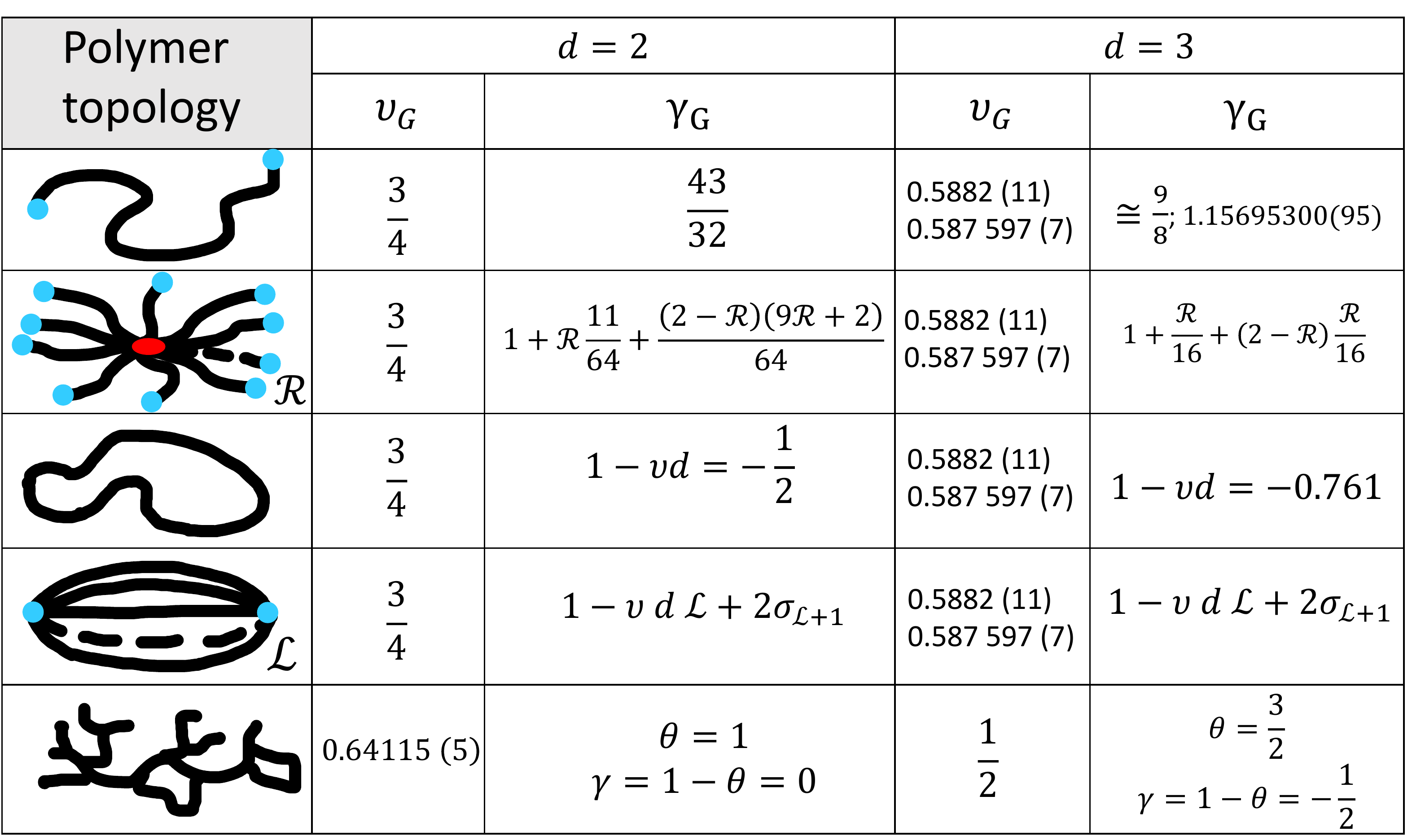}\\
\caption{Metric and entropic exponents of different polymer network
  topologies (i.e. homeomorphism types) in $d=2$ and $d=3$ under good
  solvent conditions. First row refers to linear self-avoiding
  polymers: values in $d=2$ are exact and were first computed via
  Coulomb gas~\cite{nienhuis1982exact}; $d=3$ values have been
  originally obtained by Wilson-Fisher expansion in dimension
  $d=\epsilon-4$ for the $\varphi^4$ field theory of the $O(n\to 0)$
  model with $\epsilon=1$~\cite{guida1998} (first number), and later
  estimated with high precision Monte Carlo
  simulations~\cite{clisby2010accurate,clisby2017scale} (second
  number).  Exponents from second to fourth row refer to star polymers
  with $\cR$ arms~\cite{duplantier1989}, ring polymers, and watermelon
  graphs with $\cL$ independent loops~\cite{duplantier1989},
  respectively. Since topology is kept fixed, the $\nu$ exponent does
  not vary with respect to the linear case. The $\gamma$ exponent is
  obtained filling proper values for $\nu$, $d$, $\mathcal{L}$, $n_L$
  in Eq.~\eqref{eq_gamma}, together with $\sigma_L = (2-L)(9L+2)/64$
  (exact, $d=2$) or $\sigma_L= \epsilon (2-L)L/16 + O(\epsilon^2)$
  ($d=3$).  For instance, with star polymers one has $\cL=0$,
  $n_1=\cR$, and $n_{\cR}=1$ (the $n_2$ monomers with functionality
  $L=2$ do not contribute as $\sigma_2=0$) -- note that the $2$-arms
  topology ($n_1=2$, $\cR=2$) corresponds to linear polymer.
  Similarly, with watermelon graphs $\cL\geq1$, $n_{\cL+1}=2$.  The
  last row refers instead to lattice animals, i.e. polymers in which
  the number of loops and branches can vary. In this case critical
  exponents in $d$ dimensions are related to the Lee-Yang edge
  singularity~\cite{parisi1981critical} through the relations
  $\nu_{d+2}=(\beta_d+1)/2$ and $\theta_{d+2}=\beta_d+2$, where
  $\beta_d$ is the exponent controlling magnetization near the edge
  singularity. As the latter is exactly solvable in $d=0$ and $d=1$
  with $\beta_0=-1$ and $\beta_1=-1/2$, respectively, one gets
  $\gamma_2=1-\theta_2=0$, $\gamma_3=1-\theta_3=-1/2$, and
  $\nu_3=3/2$.  The exact expression for $\nu_{d+2}$ however breaks
  down with $d=0$ and for $\nu_2$ one has to rely on numerical
  estimates~\cite{jensen2001enumerations}, such as the one reported in
  table. Note that with $d\ge 3$ exponents $\gamma$ and $\nu$ are not
  independent but follow the relation $\gamma=1-(d-2)\nu$.  }
\label{fig_exponents}
\end{figure}
\end{center}

Consider the grand canonical description of an isolated polymer
in solution in contact with a monomer chemostat. The size of the
polymer $N$ is a random variable 
and to the event $N=n\in\mathbb{N}$ is associated
an equilibrium distribution $P^\star_N(n)$  
determined by the monomer fugacity $z$. Close to criticality
\mbox{$z\to z_{\mathrm{c}}^-$} the partition function asymptotically
behaves as
\begin{equation}
  Z_{\mathrm{gc}}(z)=\sum_{n}(\mu_{\mathrm{c}}\,z)^n\;n^{\gamma-1}
  \sim
  \left\{ \begin{array}{ll}
    (1-z/z_{\mathrm{c}})^{-\gamma} & (\gamma>0)\\
    -\ln(1-z/z_{\mathrm{c}}) & (\gamma=0)\,,\\
    \textrm{finite} & (\gamma<0)
  \end{array} \right.
\end{equation}
where $\mu_{\mathrm{c}}$ is the (model-dependent) connective constant
and $z_{\mathrm{c}}=\mu_{\mathrm{c}}^{-1}$.  The entropic
exponent $\gamma$ is specified by the space dimension $d$ and by the
topology of the polymeric structure (homeomorphism type of
the underlying graph); together
with the metric exponent $\nu>0$ it identifies the universality class of
the critical behavior.  For the wide class of
polymer networks in good solvent conditions, with any prescribed fixed
topology $\mathcal{G}$ made of chains of equal lengths, this
exponent is known thanks to the mapping with the magnetic $O(n\to 0)$
model~\cite{deGennes1979} through the relation~\cite{duplantier1989}
\begin{equation}
  \gamma=\gamma_{\mathcal{G}}
  =1-\nu\,d\,\mathcal{L}+\sum_{L\geq1}n_L\,\sigma_L\,,
  \label{eq_gamma}
\end{equation}
where $\mathcal{L}$ is the number of physical loops (or cyclomatic
number) in the polymer network, $n_L$ the number of vertices with
functionality $L$, and 
$\sigma_L$ the associated scaling dimension (see
Fig.~\ref{fig_exponents} for examples and further details).  Also in
the case when monomers functionality is free to fluctuate as in
lattice animals and
trees~\cite{lubensky1979statistics,vanderzande1998}, the $\gamma$
exponent can be exactly computed by relating the critical behavior of
these systems to the Yang-Lee singularity of an Ising model in $d-2$
dimensions~\cite{parisi1981critical,brydges2003branched} (again, more
details in Fig.~\ref{fig_exponents}).
The metric exponent characterizes the
large $N$ behavior of the average square end-to-end distance of large
polymer chains: \mbox{$R^2\sim N^{2\nu}$}. Unlike the entropic
one, its value does not depend on topology (if fixed) but on
the dimension of the embedding space (see Fig.~\ref{fig_exponents}).
Note that both exponents 
can further depend on the polymer being in different equilibrium phases 
such as those triggered by monomer-monomer attractions (coil to globule transition)
or by effective interactions with impenetrable surfaces (adsorption transition)~\cite{deGennes1972,vanderzande1998}.
 
\onecolumngrid
\twocolumngrid

The position of the polymer CM 
$\boldsymbol{R}_{\mathrm{CM}}(t)=(X_{\mathrm{CM}}(t),Y_{\mathrm{CM}}(t),Z_{\mathrm{CM}}(t))$
 undergoes a Brownian motion
described by~\cite{notesmoluchowski}
\begin{equation}
  \mathrm{d}\boldsymbol{R}_{\mathrm{CM}}(t)=\sqrt{2D(N(t))}
  \,\mathrm{d}\boldsymbol{B}(\mathrm{d}t)\,,
  \label{eq_smoluchowski}
\end{equation}
where $\boldsymbol{B}(t)$ is a Wiener process (Brownian motion).  In
view of the Stokes-Einstein relation~\cite{Doi1992,metzler2021,note_stokes_einstein} 
\begin{equation}
  D(N)\sim D_0/N^\nu\,,
  \label{eq_nu}
\end{equation}
with $D_0$ specific of the polymer subunits.
Under the present assumptions $D$ fluctuates with the polymer
size: we see that $\boldsymbol{R}_{\mathrm{CM}}(t)$ becomes
thus a subordinated stochastic process~\cite{Feller1968} conditioned
by the history
$[n(t)]\equiv\{n(t')\in\mathbb{N}\;|\;0\leq t'\leq t\}$ of the
polymer size. It is convenient to reparametrize the diffusion path in
terms of the coordinate $s\geq0$,
$\mathrm{d}s=2\,D(n(t))\,\mathrm{d}t$, corresponding to the
realization of the stochastic process
\begin{equation}
  S(t)\equiv2\int_0^t\mathrm{d}t'\,D(N(t'))
  =2D_0\int_0^t\mathrm{d}t'\,N^{-\nu}(t')\,.
\end{equation}
By using  the subordination formula~\cite{Feller1968,bochner2020harmonic}
\begin{equation}
  p_{\boldsymbol{R}_{\mathrm{CM}}}(\boldsymbol{r},t|n_0;\boldsymbol{0})
  =\int_0^\infty\mathrm{d}s
  \,\dfrac{\mathrm{e}^{-\frac{\boldsymbol{r}^2}{2s}}}{(2\pi\,s)^{3/2}}
  \;p_S(s,t|n_0)\,,
  \label{eq_subordination}
\end{equation}
where
$p_{\boldsymbol{R}_{\mathrm{CM}}}(\boldsymbol{r},t|n_0;\boldsymbol{0})$
is the CM conditional PDF given the
initial condition
$p_{\boldsymbol{R}_{\mathrm{CM}}}(\boldsymbol{r},0|n_0;\boldsymbol{0})=\delta_{n,n_0}\,\delta(\boldsymbol{r})$,
moments of the subordinated process are straightforwardly connected to
those of the subordinator. For instance, assuming an equilibrium
distribution $P^\star_N(n_0)$  for the initial size, 
\begin{equation}
  \mathbb{E}[X_{\mathrm{CM}}^2(t)]
  =\mathbb{E}[S(t)],
  \quad
  \mathbb{E}[X_{\mathrm{CM}}^4(t)]
  =3\,\mathbb{E}[S^2(t)]\,.
  \label{eq_averages}
\end{equation}
We already appreciate the influence of $\gamma$ and $\nu$ in these
moments: while the latter enters in the definition of $S(t)$, the
former characterizes $P^\star_N(n)$. 

Importantly, the equilibrium distribution $P^\star_N(n)$ can be
related to a simple master equation describing the
polymerization/depolymerization process occurring as monomers add and
detach to the polymer in the grand canonical ensemble.
First of all we observe that if $n_{\mathrm{min}}$ is the minimal
polymer size, through the change of variable
$n\mapsto n-(n_{\mathrm{min}}-1)$ we can always associate the support
$1\leq n<\infty$ to $P^\star_N(n)$
without altering the asymptotic behavior in Eq.~\eqref{eq_nu}.
Regard then the (forward) master equation 
\begin{eqnarray}
  &&\partial_t P_N(n,t|n_0)
  =
  \mu\,P_N(n+1,t|n_0)
  \quad\;\;(n>1)
  \label{eq_master}\\
  &&\;\;\;
  +\lambda(n-1)\,P_N(n-1,t|n_0)
  -(\mu+\lambda(n))\,P_N(n,t|n_0)\,
  \nonumber\\
  &&
  \partial_t P_N(1,t|n_0)=
  \mu\,P_N(2,t|n_0)
  -\lambda(1)\,P_N(1,t|n_0)\,.
  \nonumber
\end{eqnarray}
Here $P_N(n,t|n_0)$ is the probability for $N=n$ at time $t\geq0$ given
$N=n_0$ at $t=0$, and $\lambda(n)$, $\mu$ are
the rates for association and dissociation, respectively.
Defining the growth factor as $g(n)\equiv \lambda(n)/\mu$, it is
straightforward to prove that stationarity is attained under detailed
balance,
$g(n)=P^\star_N(n+1)/P^\star_N(n)=\mu_{\mathrm{c}}z\,[(n+1)/n]^{\gamma-1}$:
this identifies the 
polymerization process, given $P^\star_N(n)$.
In \textit{chain polymerization}~\cite{odian2004},
while it is natural to consider dissociation to be independent of the
polymer size, aggregation is instead influenced by the ratio of the
number of available configurations at sizes $n+1$ and $n$. 
This is the reason of the size-dependency $\lambda(n)$ assumed here, which
is conveyed by the entropic correction $\propto n^{\gamma-1}$ outside the
mean-field limit ($\gamma\neq1$).  
Note that the rate $\mu$
remains a free parameter which may rescale Eq.~\eqref{eq_master},
thus determining the time scale $\tau$ for 
the autocorrelation of $N(t)$. This is particularly apparent in 
the mean-field case
($\gamma=1$), where an elegant connection
with the $M/M/1$ model (Markovian interarrival
times/Markovian service times/1 server) in queuing
theory~\cite{Jain2007} allows to extend the identification even
outside criticality and to analytically solve both the equilibrium and
the out-of-equilibrium behavior~\cite{nampoothiri2021}.
The asymptotic behavior for small and large time of $P_N(n,t|n_0)$ is
\mbox{$P_N(n,t|n_0)\substack{\sim\\t\ll\tau}\delta_{n,n_0}$},
\mbox{$P_N(n,t|n_0)\substack{\sim\\t\gg\tau}P^\star_N(n)$},
respectively.

The equilibrium size distribution is directly deduced from the grand
canonical partition function (generating function). 
Close to the critical point
$z_{\mathrm{c}}=1/\mu_{\mathrm{c}}$ we may neglect regular
contributions, and, remembering the definition of
polylogarithm functions, $\mathrm{Li}_s(z)\equiv\sum_{n=1}^\infty
z^n/n^s$ (which are finite in $z=1$ if $s>1$), we have
\begin{eqnarray}
  P^\star_N(n)&=&\dfrac{(z/z_{\mathrm{c}})^n
  }{n^{1-\gamma}\;\mathrm{Li}_{1-\gamma}(z/z_{\mathrm{c}})}
  \nonumber\\
  &\sim&
  \left\{ \begin{array}{ll}
    \dfrac{
      (1-z/z_{\mathrm{c}})^{\gamma}\;(z/z_{\mathrm{c}})^n
    }{n^{1-\gamma}\;\Gamma(\gamma)}
    & (\gamma>0)\\
    -\dfrac{(z/z_{\mathrm{c}})^n}{n\,\ln(1-z/z_{\mathrm{c}})} & (\gamma=0)\,.\\
    \dfrac{(z/z_{\mathrm{c}})^n}{n^{1-\gamma}\,\mathrm{Li}_{1-\gamma}(z/z_{\mathrm{c}})} & (\gamma<0)
  \end{array} \right.
  \label{eq_distribution}
\end{eqnarray}

We are now in a position to evaluate expected values in
Eq.~\eqref{eq_averages}.
Let us primarily note that, with equilibrium initial conditions
$P^\star_N(n_0)$,
\begin{equation}
  \begin{array}{ll}
    \mathbb{E}[S(t)]
    &\sim2D_0\displaystyle\int_0^t\mathrm{d}t'\sum_{n,n_0}
    \dfrac{P_N(n,t'|n_0)\,P^\star_N(n_0)}{n^{\nu}}\\
    &=2D_0\,t\,\mathbb{E}[N^{-\nu}]\,,
  \end{array}
\end{equation}
where we have
used the stationarity of $P^\star_N(n)$.
Together with Eq.~\eqref{eq_averages}, this proves
the Brownian character of the CM diffusion in equilibrium.
Transients may display either sub- or super-diffusive stages, depending
on the specific initial
condition~\cite{nampoothiri2021}.
Using the asymptotic expressions for $P_N(n,t|n_0)$, we analogously
find
\begin{equation}
  \mathbb{E}[S^2(t)]
  \sim
  \left\{ \begin{array}{ll}
    (2D_0\,t)^2\,\mathbb{E}[N^{-2\nu}]
    & (t\ll\tau)\\
    (2D_0\,t)^2\,(\mathbb{E}[N^{-\nu}])^2
    & (t\gg\tau)\\
  \end{array} \right.,
\end{equation}
which implies, for the CM kurtosis,
\begin{equation}
  \kappa_{\mathrm{CM}}(t)=\dfrac{3\,\mathbb{E}[S^2(t)]}{(\mathbb{E}[S(t)])^2}
  \sim
  \left\{ \begin{array}{ll}
    3\,\dfrac{\mathbb{E}[N^{-2\nu}]}{(\mathbb{E}[N^{-\nu}])^2}
    & (t\ll\tau)\\
    3\;\textrm{(Gaussian)}
    & (t\gg\tau)\\
  \end{array} \right..
  \label{eq_kurtosis}
\end{equation}
Eq.~\eqref{eq_kurtosis} shows that, while the kurtosis is potentially
different from $3$ for time within the scale $\tau$, it crosses over
to the Gaussian value at larger time.  This is specifically what is
observed in many
experiments~\cite{wang2009,wang2012,jeon2016,cherstvy2018,li2019,cuetos2018}
and also obtained in various mesoscopic
models~\cite{chechkin2017,jain2017,tyagi2017,sposini2018,wang2020unexpected,miotto2021length}.

\begin{table}
  \begin{tabular}{|c|c|}
    \hline
    \multicolumn{2}{|c|}{$\gamma>0$} \\
    \hline
    $\nu$ & $\kappa_{\mathrm{CM}}(t)\substack{\sim\\t\ll\tau}$ \\
    \hline
    $0<\nu<\gamma/2$
    &
    $3\dfrac{\Gamma(\gamma)\,\Gamma(\gamma-2\nu)}{[\Gamma(\gamma-\nu)]^2}$
    (finite)\\
    \hline
    $0<\nu=\gamma/2$
    &
    $3\dfrac{\Gamma(\gamma)}{[\Gamma(\gamma/2)]^2}[-\ln(1-z/z_{\mathrm{c}})]$\\
    \hline
    $\gamma/2<\nu<\gamma$
    &
    $3\dfrac{\Gamma(\gamma)
      \,\mathrm{Li}_{1-\gamma+2\nu}(z/z_{\mathrm{c}})
    }{[\Gamma(\gamma-\nu)]^2}
    \dfrac{1}{(1-z/z_{\mathrm{c}})^{2\nu-\gamma}}$\\
    \hline
    $\nu=\gamma$
    &
    $3\dfrac{\Gamma(\gamma)\,
      \,\mathrm{Li}_{\gamma+1}(z/z_{\mathrm{c}})
    }{(1-z/z_{\mathrm{c}})^{\gamma}\;[-\ln(1-z/z_{\mathrm{c}})]^2}$\\
    \hline
    $\nu>\gamma$
    &
    $3\dfrac{\Gamma(\gamma)\,
      \,\mathrm{Li}_{1-\gamma+2\nu}(z/z_{\mathrm{c}})
    }{[\mathrm{Li}_{1-\gamma+\nu}(z/z_{\mathrm{c}})]^2}
    \dfrac{1}{(1-z/z_{\mathrm{c}})^{\gamma}}$\\
    \hline
    \hline
    \multicolumn{2}{|c|}{$\gamma=0$} \\
    \hline
    $\nu>0$
    &
    $3\dfrac{\mathrm{Li}_{1+2\nu}(z/z_{\mathrm{c}})
    }{[\mathrm{Li}_{1+\nu}(z/z_{\mathrm{c}})]^2}
    [-\ln(1-z/z_{\mathrm{c}})]$\\
    \hline
    \hline
    \multicolumn{2}{|c|}{$\gamma<0$} \\
    \hline
    $\nu>0$
    &
    $3\,Z_{\mathrm{gc}}(z)\dfrac{\mathrm{Li}_{1-\gamma+2\nu}(z/z_{\mathrm{c}})
    }{[\mathrm{Li}_{1-\gamma+\nu}(z/z_{\mathrm{c}})]^2}$ (finite)\\
    \hline
  \end{tabular}
  \caption{Behavior of the initial CM kurtosis.}
  \label{tab_kurtosis}
\end{table}

To evaluate the non-Gaussianity of the CM diffusion in terms of
$\kappa_{\mathrm{CM}}$ during the early
stages, averages in
Eq.~\eqref{eq_kurtosis} 
must be calculated according to Eq.~\eqref{eq_distribution}. Once
more, this invokes the known behavior of the polylogarithm function; it
also highlights the interplay between exponents $\gamma$, $\nu$ in
establishing the dynamic response. We have
\begin{equation}
  \kappa_{\mathrm{CM}}(t)\substack{\sim\\t\ll\tau}
  \dfrac{3}{\mathrm{Li}_{1-\gamma}(z/z_{\mathrm{c}})}
  \dfrac{\mathrm{Li}_{1-\gamma+2\nu}(z/z_{\mathrm{c}})}
        {[\mathrm{Li}_{1-\gamma+\mu}(z/z_{\mathrm{c}})]^2}\,.
\end{equation}
Table~\ref{tab_kurtosis} wraps up the initial kurtosis behavior, which
includes power-law divergency (possibly with log-corrections),
logarithmic divergency, or even finiteness.
The shape of the initial non-Gaussian PDF for the polymer CM is
conveniently studied by switching to the unit-variance
dimensionless variable
$\overline{X}_{\mathrm{CM}}(t)\equiv X_{\mathrm{CM}}(t)/\sqrt{\mathbb{E}[X^2(t)]}$.
From Eq.~\eqref{eq_subordination}, as $t\to0^+$, we have
\begin{equation}
  p_{\overline{X}}(x,0^+)
  \sim\sum_{n=1}^\infty P^\star_N(n)
  \,\dfrac{\mathrm{e}^{-\frac{\mathbb{E}[N^{-\nu}]\,x^2}{2\,n^{-\nu}}}
  }{
    \sqrt{2\pi\,\frac{n^{-\nu}}{\mathbb{E}[N^{-\nu}]}}
  }\,,
\end{equation}
which only depends on $z/z_{\mathrm{c}}$, $\gamma$, and $\nu$.
At large $|x|$ the PDF is
asymptotic to the Gaussian cutoff
$\sim\mathrm{e}^{-\mathbb{E}[N^{-\nu}]\,x^2/2}$, and as
$z/z_{\mathrm{c}}\to1$ 
this cutoff is pushed towards $|x|\to\infty$, since
$\mathbb{E}[N^{-\nu}]\to0$. 

\begin{figure}[t]
\includegraphics[width=0.9\columnwidth]{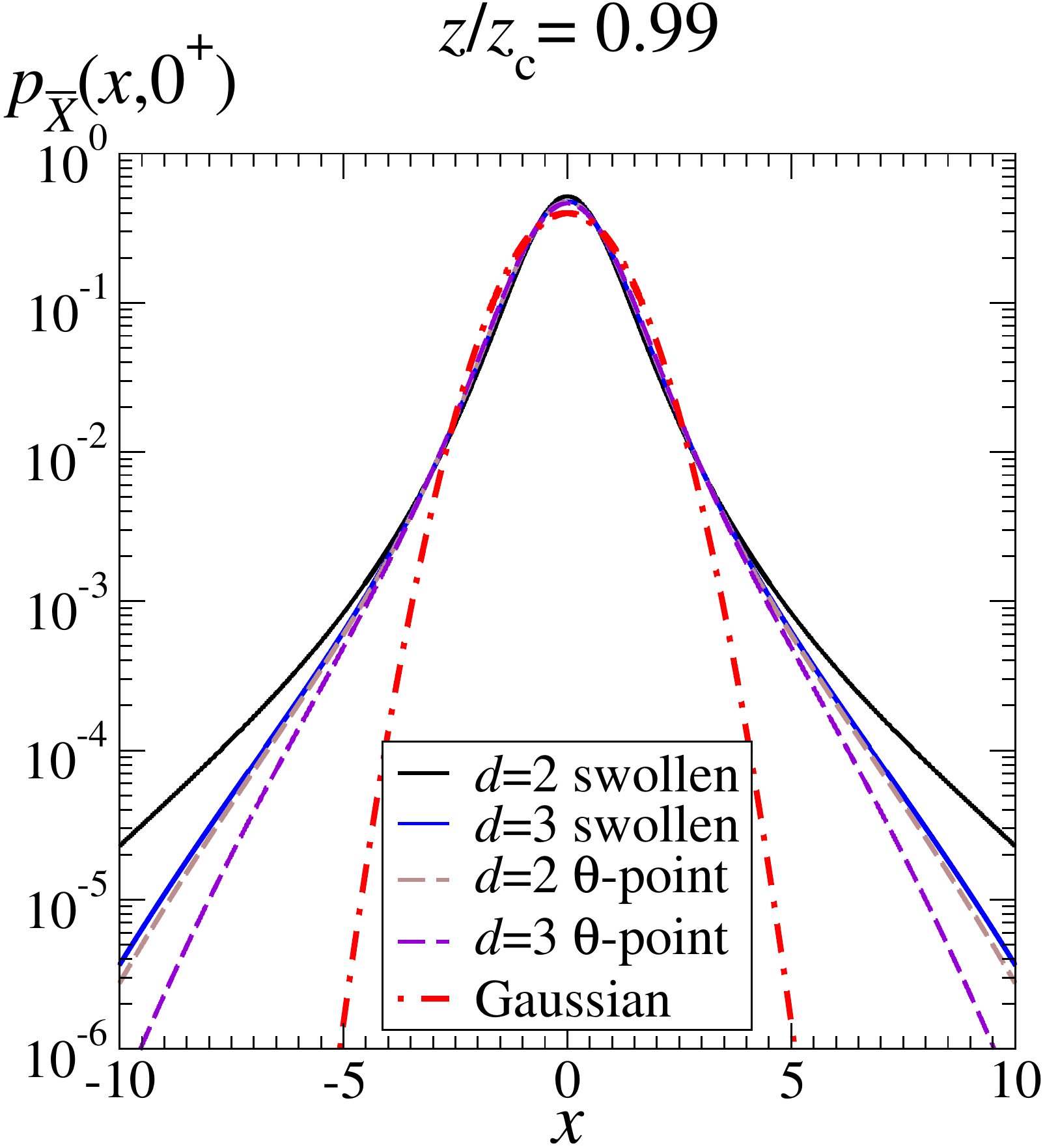}\\
\caption{Unit-variance initial $x$-PDF for the CM of a linear polymer
  close to criticality.
  For comparison purposes, a unit-variance Gaussian
  PDF is also plotted in red dash-dotted line.}
\label{fig_initial_pdf_linear}
\end{figure}

\begin{figure}[t]
\includegraphics[width=0.9\columnwidth]{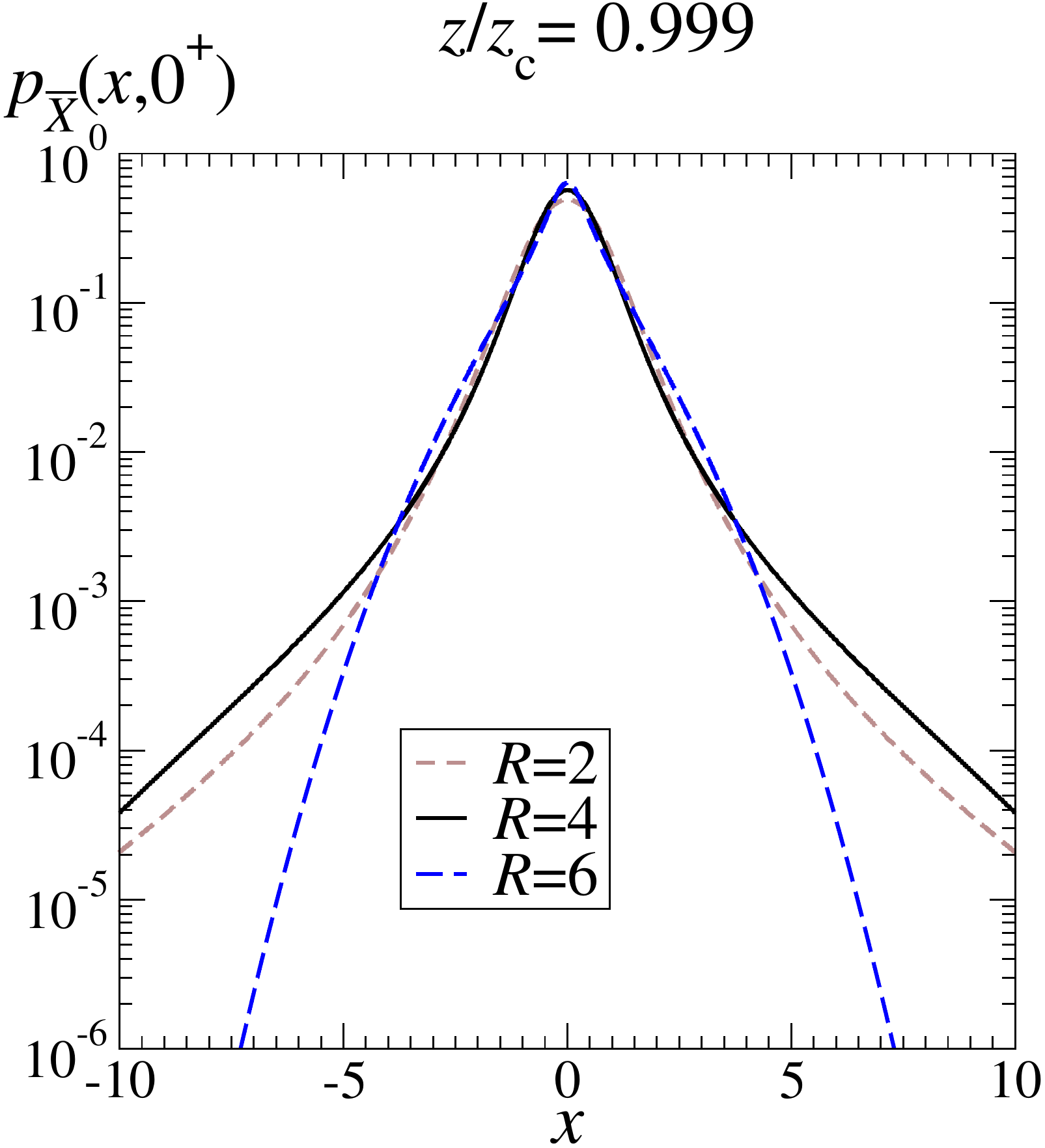}\\
\caption{Unit-variance initial $x$-PDF for the CM of a star polymer
  in $d=3$ with different numbers of arms $\mathcal{R}$ close to
  criticality.  
}
\label{fig_initial_pdf_star}
\end{figure}

It is now interesting to discuss peculiar initial dynamical responses of polymers 
with different topologies, as $z\to z_{\mathrm{c}}$.\\
\textit{Linear polymers}. In this case the condition $\gamma/2 < \nu
<\gamma$ is satisfied both in $d=2$ and $d=3$ (see first row
in Fig~\ref{fig_exponents}) and the kurtosis diverges with exponents 
$5/32$ and $\simeq 0.012$, respectively --
cf. Fig.~\ref{fig_initial_pdf_linear}.  \\ 
\textit{$\mathcal{R}$-arms star polymers}.  In $d=2$ the kurtosis
diverges if $\mathcal{R}\leq4$, with exponent $2\nu-\gamma$
($\mathcal{R}=2,3$) or $\gamma$ ($\mathcal{R}=4$).  In $d=3$ the
kurtosis diverges if $\mathcal{R}\leq5$, with exponent $2\nu-\gamma$
($\mathcal{R}=2,3,4$) or $\gamma$ ($\mathcal{R}=5$) --
cf. Fig~\ref{fig_initial_pdf_star}. \\ 
\textit{Rings and watermelon networks}.  Since $\gamma < 0$ both in $d=2$ and $d=3$,
$\kappa_{\mathrm{CM}}$ does not diverge. \\
\textit{Branched polymers (lattice animals)}. In $d=2$ $\gamma=0$ and the kurtosis diverges
logarithmically, independently on the value of $\nu$. Instead, in $d=3$ $\gamma<0$, implying a finite value of
$\kappa_{CM}$ also at the critical point.\\

So far we have considered polymers whose equilibrium properties are
dominated by monomer-solvent attraction (swollen phase).  On the other
hand, by varying solvent conditions polymers may undergo a
thermodynamic transition from swollen (good solvent) to globular or
compact phase (poor solvent).  The transition occurs at a well defined
critical phase known as $\Theta$-point, a genuine $O(n\to 0)$
tricritical point governing an equilibrium phase characterized by its
own critical exponents $\gamma_{\Theta}$ and
$\nu_{\Theta}$~\cite{deGennes1972,vanderzande1998}. For
instance, linear polymers at the $\Theta$-point in $d=2$ have
$\nu_{\Theta}=4/7$ and $\gamma_{\Theta}=8/7$\cite{duplantier1987exact,duplantier1988polymer,seno1988theta,vanderzande1991percolation}, whereas in $d=3$ the
mean-field values $\nu_{\Theta}=1/2$ and $\gamma_{\Theta}=1$ are
expected~\cite{deGennes1979,duplantier1982lagrangian}.
Hence, in both dimensions $\nu=\gamma/2$.  This remarkable relation
has the important consequence that as $z\to z_c$ the initial kurtosis
diverges logarithmically for linear polymers at the $\Theta$-point,
irrespective of the dimension. This result suggests that a change in
the quality of the solvent driving dilute linear polymers close to
$\Theta$-point, concomitantly mitigates the non-Gaussianity of the CM
diffusion from power-law to logarithmic divergence of $\kappa_{CM}$.
Fig.~\ref{fig_initial_pdf_linear} displays the associated PDFs.

Finally, since the $\nu$ and $\gamma$ exponents depend on the
embedding dimension $d$ of the system, transitions between phases with
different effective dimension may also alter
the non-Gaussianity of the initial CM diffusion.  An example is the well
studied~\cite{deGennes1972,vanderzande1998}
adsorption transition from the $d=3$ polymer swollen phase to
the adsorbed ($d=2$) swollen
phase. This is triggered by
effective attractive interactions between monomers
and an impenetrable surface. With a non negligible mobility of the
polymer at the surface, the adsorption transition of linear polymers
increases the exponent of the power law divergence of
$\kappa_{CM}$ from $0.012$ ($d=3$) to $5/32$ ($d=2$).

We have analytically shown that the polymer critical state is the
hallmark behind the non-Gaussian behavior of its CM.
To each universality class, identified by the entropic and metric
exponents $\gamma$ and $\nu$, corresponds a specific Brownian
non-Gaussian diffusion of the polymer CM which crosses then over to
ordinary Brownian motion above the polymerization autocorrelation time
scale. This finding offers novel perspectives in stochastic modeling,
as the anomalous stochastic process is not obtained here via a mesoscopic ansatz, but rather 
as a natural consequence of a microscopic foundation  which can be worked out in all details and bridges the 
universal behavior of polymer systems at equilibrium with their short time anomalous dynamical response. 
The background we have evoked (different polymer architectures,
$\Theta$- and adsorption transitions) is commonly operated in
polymer experiments; this implies the
exposed anomalous dynamics to be potentially 
triggered and highlighted in a variety of chemostatted experimental conditions.

\section*{Acknowledgments}
We acknowledge insightful discussions with R. Metzler. This work has
been partially supported by the University of Padova BIRD191017
project ``Topological statistical dynamics''.

\bibliography{draft_bibliography}

\begin{thebibliography}{59}
\bibitem [{not({\natexlab{a}})}]{notesmoluchowski}%
  \BibitemOpen
  \href@noop {} {\bibfield  {journal} {\bibinfo  {journal} {This equation
  applies to the Smoluchowski time scale; the latter is appropriate for the
  description of the polyer dynamics at any sizes $N$ varying from single
  monomer to large colloids. Indeed, in the time needed to loose memory of
  inertial effects, the travelled distance with respect to the polymer radius
  $a$ is given by $\sqrt{3\,m\,k_{\mathrm{B}}T}/(6\pi\,\eta\,a^2)\,.$ In water
  at room temperature, this ratio varies from $10^{-3}$ for single nucleotides
  or amino acids to $10^{-5}$ for large colloids}\ }
  ({\natexlab{a}})}\BibitemShut {NoStop}%
\bibitem [{not({\natexlab{b}})}]{note_stokes_einstein}%
  \BibitemOpen
  \href@noop {} {\bibfield  {journal} {\bibinfo  {journal} {The value for the
  diffusion coefficient $D$ to be filled in Eq.~\eqref{eq_smoluchowski} is the
  one coming from the hydrodynamic radius of the polymer
  $R$~\cite{rubinstein2003}. Close to criticality, the latter is determined by
  the metric exponent $\nu$ as $R\sim N^{\nu}$. So, Eq.~\eqref{eq_smoluchowski}
  can be considered with both the exact values for the exponent $\nu$ and their
  approximations, like, e.g., the Flory one~\cite{Doi1992,rubinstein2003}}\ }
  ({\natexlab{b}})}\BibitemShut {NoStop}%
\end{thebibliography}

\end{document}